\documentclass{mn2e}
\usepackage{graphicx}

\def\ltsima{$\; \buildrel < \over \sim \;$}
\def\lsim{\lower.5ex\hbox{\ltsima}}
\def\gtsima{$\; \buildrel > \over \sim \;$}
\def\gsim{\lower.5ex\hbox{\gtsima}}

\newcommand{\be}{\begin{equation}}
\newcommand{\en}{\end{equation}}

\def\cmdue {\rm \ cm^{-2}}

\begin{document}
   \title[Probing the very high redshift Universe with GRB]{Probing the very high redshift Universe with 
   Gamma--ray Bursts: prospects for observations with future X--ray instruments}

\author[S. Campana et al.]{S. Campana$^{1,}$\thanks{E-mail: sergio.campana@brera.inaf.it}, R. Salvaterra$^2$, G. Tagliaferri$^1$, 
C. Kouveliotou$^3$, J. Grindlay$^4$\\
$^1$ INAF-Osservatorio Astronomico di Brera, Via Bianchi 46, I--23807, Merate (Lc), Italy\\
$^2$ Universit\`a dell'Insubria, Via Valleggio 11, 22100 Como, Italy\\
$^3$ NASA Marshall Space Flight Center, Huntsville, Alabama 35805, USA\\
$^4$ Harvard-Smithsonian Center for Astrophysics, Cambridge, Massachusetts 02138, USA
}

\maketitle

\begin{abstract}
Gamma-Ray Bursts (GRBs)  are the most violent explosions in the Universe. Long duration GRBs 
are associated with the collapse of massive stars, rivaling their host galaxies in luminosity.
The discovery of the most distant spectroscopically confirmed object in the Universe, GRB 090423, opened 
a new window on the high redshift Universe, making it possible to study the cosmic Ôre-ionizationÕ epoch and the 
preceding dark ages, as well as the generation of the first stars
(Population III) using GRBs. Obviously this enables a  
wealth of new studies using the near infrared (nIR) characteristics of GRB afterglows. 
Here we explore a different path, focusing on the next generation of X--ray missions with large area focusing 
telescopes and fast repointing capabilities. 
We found that X--ray data can complement nIR observations and for the brightest GRBs can provide an accurate and 
independent redshift determination. Metallicity studies can also be carried out profitably once the redshift is known.
Finally we discuss observational signatures of GRBs arising from Population III stars in the X--ray band.
\end{abstract}


\section{Introduction}

The discovery of GRB090423 at redshift $z\sim 8.2$ (Salvaterra et al. 2009; Tanvir et al. 2009) 
has shown that Gamma-Ray Bursts (GRBs) can directly probe the very distant Universe.
GRBs are so bright that they can `illuminate', for a short time, regions of distant galaxies
otherwise unobservable. The optical afterglow of one of the most distant objects ever observed 
(GRB050904 at redshift $z=6.3$, Kawai et al. 2006) was observed
with a 25 cm telescope (B\"oer et al. 2006) and could have been detected up to $z\sim 10$ 
with the Swift BAT (Cusumano et al. 2007). The optical afterglow of GRB080319B could have been observed 
with the naked eye (Racusin et al. 2008) and GRB080319B could also have been detected by Swift-BAT up 
to $z\sim 10.7$ and up to $z\sim 32$ at the nominal EXIST sensitivity (Bloom et al. 2009).
The use of GRBs as flashlights allows us to cover a large number of astrophysical topics, from
chemical evolution and reionization of the Universe to dust formation and composition.

Several studies indicate that (long duration) GRBs are associated with the death of massive stars 
(e.g. Bloom \& Woosley 2006) and their explosion sites are concentrated in the very brightest regions 
of their host galaxies more than the core-collapse supernovae (Fruchter et al. 2006; Svensson et al. 2010). 
The high level of X--ray absorption often observed suggests that GRBs occur in the densest part of their host galaxy 
(Campana et al. 2009). This implies that the GRB light in distant
galaxies shines through a gas medium that is very different from the
typical interstellar medium (ISM) observed with traditional tools,
such as quasar studies. 

The ISM in distant galaxies is usually probed by using bright higher-redshift quasi-stellar objects (quasars) 
which are used to illuminate the intervening objects. The optical absorption lines superposed on the quasar 
spectra were used to explore the chemical enrichment in the Universe (Wolfe, Gawiser \& Prochaska 2005).
However, the investigation of the high redshift ISM backlit by (bright) quasars is biased since,
for geometrical reasons, quasars probe mainly (intervening) galaxy halos, rather than galaxy bulges or disks, and for
observational reasons since quasars are usually too faint to shine through the bulge of galaxy. In addition, since
large impact parameters are guaranteed by massive galaxies, it is not clear whether these intervening galaxies
are truly representative of the whole high-redshift population. 

GRB afterglows provide an independent tool for studying the ISM of high-redshift galaxies and practically 
the only one able to probe the densest parts, star forming regions.
High-resolution optical spectroscopic observations of GRB afterglows allow us to probe the relative abundances 
of different elements highlighting the physical state of the gas, its chemical enrichment and dust content, and the
nucleosynthesis processes in the GRB star-forming environment (Savaglio 2010).
Despite the (still) poor number statistics with respect to quasar studies, a comparison between the metallicities 
obtained by GRBs and intervening galaxies along the quasars line of sight shows that the former tend to be 
more metal-rich (on average by a factor of $\gsim 10$ at high redshift, e.g. Savaglio 2006) and to have a flatter 
or no evolution with redshift, at variance with galaxies probed by more distant quasars. 
In particular, GRB studies revealed a new picture of the gas properties in the high redshift Universe, with large metal 
column densities never observed before. A metallicity of $\sim 0.1\,Z_\odot$ has been for instance estimated from a S 
line in GRB050904 (Totani et al. 2006).

\begin{table*}
\caption{Swift spectral observations of the three GRBs at $z>6$.}
\begin{tabular}{cccccccccc}
\hline
GRB    &Mode&Start&Redshift& Galactic $N_H$      & Intrinsic $N_H$     &  Power-law           & Duration & Fluence$_{0.3-10}^*$ \\
       &    & (s) &        &($10^{20}$ cm$^{-2}$)&($10^{22}$ cm$^{-2}$)& photon index         & (ks)     & (erg cm$^{-2}$)   \\
\hline
050904 & WT & 336 & 6.29   & 3.5                 &$3.8^{+1.0}_{-0.9}$  &$1.35^{+0.03}_{-0.03}$& 0.4      &  $5.8\times 10^{-7}$ \\
050904 & PC &16843& 6.29   & 3.5                 &$2.6^{+1.0}_{-0.9}$  &$1.91^{+0.05}_{-0.05}$& 22.3   &  $2.9\times 10^{-7}$ \\
080913 & PC & 103 & 6.7    & 3.2                 &$3.1^{+5.5}_{-3.1}$  &$1.76^{+0.22}_{-0.21}$& 28.2     & $1.5\times 10^{-8}$ \\
090423 & PC & 76  & 8.1    & 2.9                 &$6.4^{+2.8}_{-2.1}$  &$1.83^{+0.10}_{-0.10}$& 14.3     & $5.1\times 10^{-8}$ \\
\hline
080319B & WT & 64 & 0.94 & 1.1                &$0.12^{+0.01}_{-0.01}$  &$1.77^{+0.01}_{-0.01}$& 1.7    & $4.1\times 10^{-6}$ \\ 
\hline
\end{tabular}

\noindent $^*$ Absorbed fluence in the observed energy range 0.3--10 keV.

\end{table*}

The circumburst medium and ISM in the host galaxy and
others along the line of sight also absorb X--ray photons.
At variance with optical, X--rays are photoelectrically absorbed by metals (Morrison \& McCammon
1983), but, as the absorption in X-rays is not sensitive to the state
of the element (at least to first order, without considering X--ray 
absorption fine structure, XAFS, which require still higher resolution), these
measurements provide a less biased view of the total absorbing column density of material.
Despite their scientific potential,  absorption studies in X--rays are underdeveloped. This is mainly 
due to the lack of collecting area (and fast slewing capabilities for the larger X--ray facilities) and lack of 
high spectral resolution comparable to optical studies. X--ray astronomy CCDs provide the analogue of 
low-resolution optical spectra; despite this, interesting results can be obtained.

As demonstrated by several studies, the X--ray absorbing column towards GRBs is usually high (Campana et al. 2006, 
2010; Stratta et al. 2005; Watson et al. 2007), which implies, at least in principle, high dust extinction. In contrast, 
GRB optical afterglows often show a lack of optical reddening. Several suggestions have been advanced to explain this
dichotomy but a clear understanding has not been reached yet (e.g. Galama \& Wijers 2001). Being very
bright and very short-lived, GRBs can alter the equilibrium state of their surrounding medium on an
observable timescale by destroying dust and photoionizing gas. These effects cause a gradual reduction of
the opacity from the X--ray band (due to progressive ionization of the metals) to the optical band (due to the
progressive destruction of the dust grains, e.g. Lazzati et al. 2001). This effect is likely too fast in the optical
band to be detected but it has been observed in X--rays in GRB050904 ($z=6.3$, Campana et
al. 2007). X--ray time-resolved spectroscopy of the burst and afterglow revealed a very high absorbing
column density at early times. The absorption decreases at later times, almost disappearing $\sim 10^4$ s after the
explosion. This behavior can be understood if the absorbing material is located very close to the GRB (a few parsecs), 
so that the flash ionization of the environment progressively strips all the electrons off the ions, decreasing the
opacity of the medium. By using a photo-ionization code tailored to this burst, Campana et al. (2007) were able to constrain the
geometric properties of the environment, finding that GRB050904 exploded within a dense metal-enriched
molecular cloud. The X--ray absorber must lie within 1--5 pc from the GRB thus probing the innermost region
in the close vicinity of the GRB explosion (a factor $>100$ closer than the best optical spectroscopic
observations). The nearby environment of GRB050904, must have a metallicity of at least several percent
solar. For other GRBs, and also for GRB050904 there is a strong discrepancy between X--ray and optical absorption, 
with optical observations not showing evidence of dust extinction. A dust mixture biased toward silicate grains
(which are more easily destroyed by UV heating) rather than the more resistant carbonaceous grains can
explain the observed discrepancy (Campana et al. 2007). 
Alternatively a very flat extinction curve that is flatter than that
for the  SMC and similar to the that observed in 
other high redshift objects (Stratta et al. 2008; Maiolino et al. 2006), could also explain the data (but see Zafar et al. 2010).

The absorption caused by metals can lead to well defined signatures in the X--ray spectra (absorption edges at
the transition energy of the most abundant elements). These edges are commonly observed in the spectra
of bright X--ray binaries in our Galaxy when observed with high (spectral) resolution gratings on board 
the current X--ray missions, XMM-Newton and Chandra. These instruments, however, have not yet observed 
very bright GRBs. The best we can do nowadays in X--rays is to observe very bright GRBs with the lower spectral resolution
provided by CCDs. Campana et al. (2008) exploited this novel technique on the burst with the
largest number of photons ever observed, GRB060218, the second closest GRB ever at $z=0.03$. For this burst we
were able to fit simultaneously the abundance of a limited number of elements (C, N and O), disentangling their
contribution from the bulk of the column density absorption. 
With this technique we obtained for the first time abundance ratios in the circumburst medium, which provide 
evidence of material highly enriched in low-atomic-number metals ejected before the GRB explosion. 
We find that, within the current scenarios of stellar evolution, only a progenitor star characterized by a fast stellar 
rotation and subsolar initial metallicity could produce such a metal enrichment in its close surroundings.

Here we explore the potential of X--ray studies with the next generation of X--ray satellites with fast repointing
capabilities to characterize the circumburst medium of high redshift GRBs.
In section 2 we describe an ideal telescope for these kind of studies and, based on the Swift view of the 
three GRBs at $z>6$, we simulate what could be observed with an
idealized $1,000$ cm$^2$ effective area X--ray telescope, 
similar to those proposed for EXIST and Xenia.
In Section 3 we discussion our results in view of the exploitation of tracing the metallicity evolution of the Universe.

\section{Simulation}

We consider here the three Swift GRBs at $z>6$: GRB050904 (Cusumano et al. 2006, 2007, Campana et al. 2007); 
GRB080913 (Greiner et al. 2009); GRB090423 (Salvaterra et al. 2009; Tanvir et al. 2009). 
The idea is to simulate the X--ray spectra of these GRBs as could be observed by an ideal telescope similar to the ones under 
studies nowadays, like EXIST (Grindlay et al 2009, 2010) or Xenia
(Kouveliotou \& Piro 2008; Piro et al. 2010). These are large X--ray facilities with an imaging  
gamma-ray burst monitor and with fast repointing capabilities. The International X--ray Observatory (IXO) does not
foresee a fast slewing capabilities and therefore, despite its large effective area, will collect less photons than the two facilities 
above due to the fast temporal decay of GRB afterglows.  
We idealize the telescope taking an effective area of 1,000 cm$^{2}$ at 1 keV (including mirror effective area, 
detector efficiency and filter transmission). We consider two possible idealized focal plane detectors: a Charge Coupled Device
(CCD) with a spectral resolution of 150 eV at 6.4 keV and a micro-calorimeter\footnote{We do not take into account detector 
efficiencies here and just assume a 1,000 cm$^2$ effective area telescope, thus requiring slightly different mirror 
assembly for the two cases.} with 2 eV constant spectral resolution.
These choices are dictated also by the need to explore how much these studies are sensitive to spectral resolution. 
The considered energy band is 0.2--10 keV.

\begin{figure}
\begin{center}
\includegraphics[width=5cm,angle=-90]{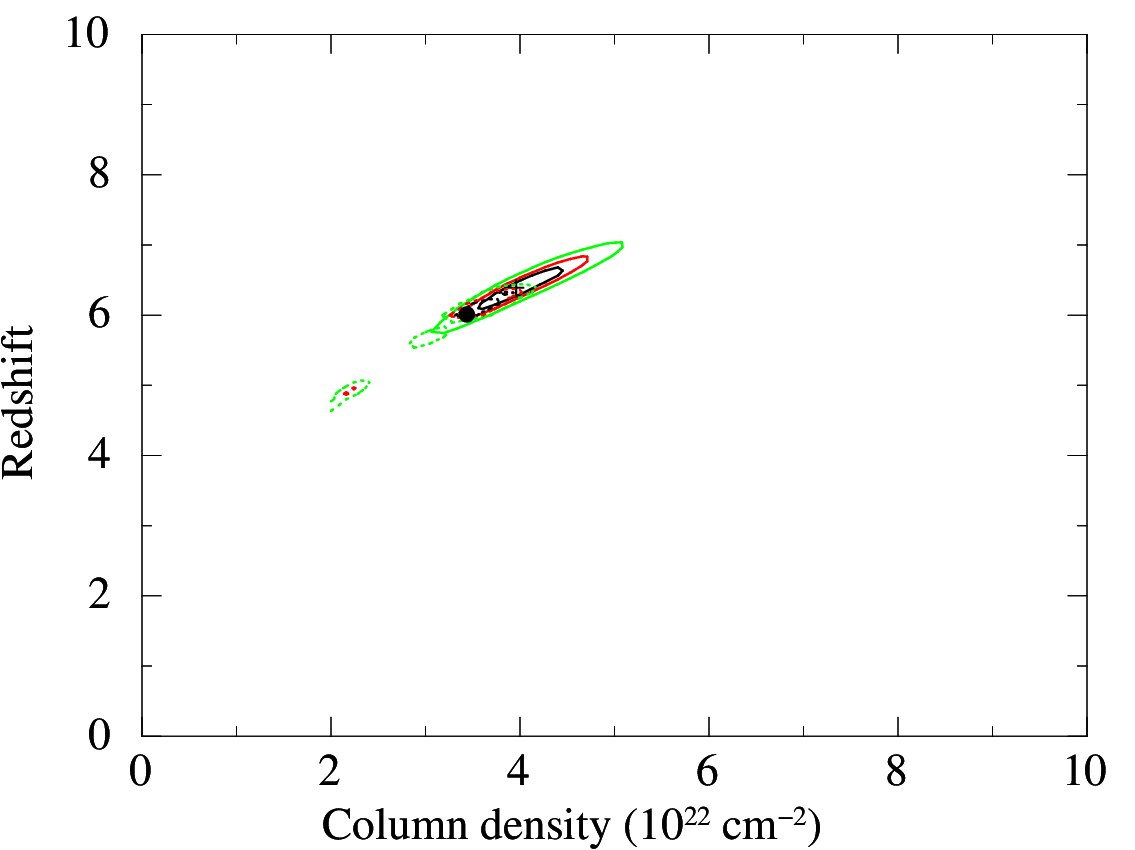}\\
\vskip 1truecm
\includegraphics[width=5cm,angle=-90]{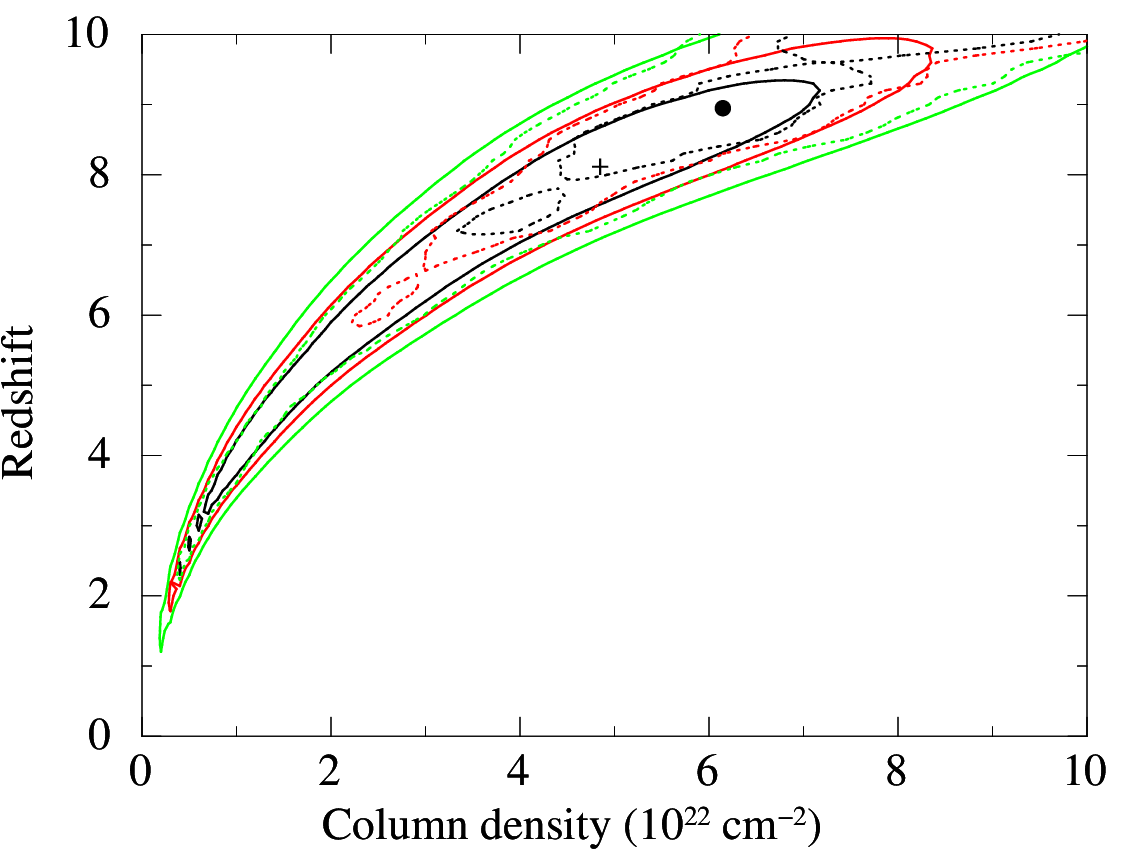}\\
\vskip 1truecm
\includegraphics[width=5cm,angle=-90]{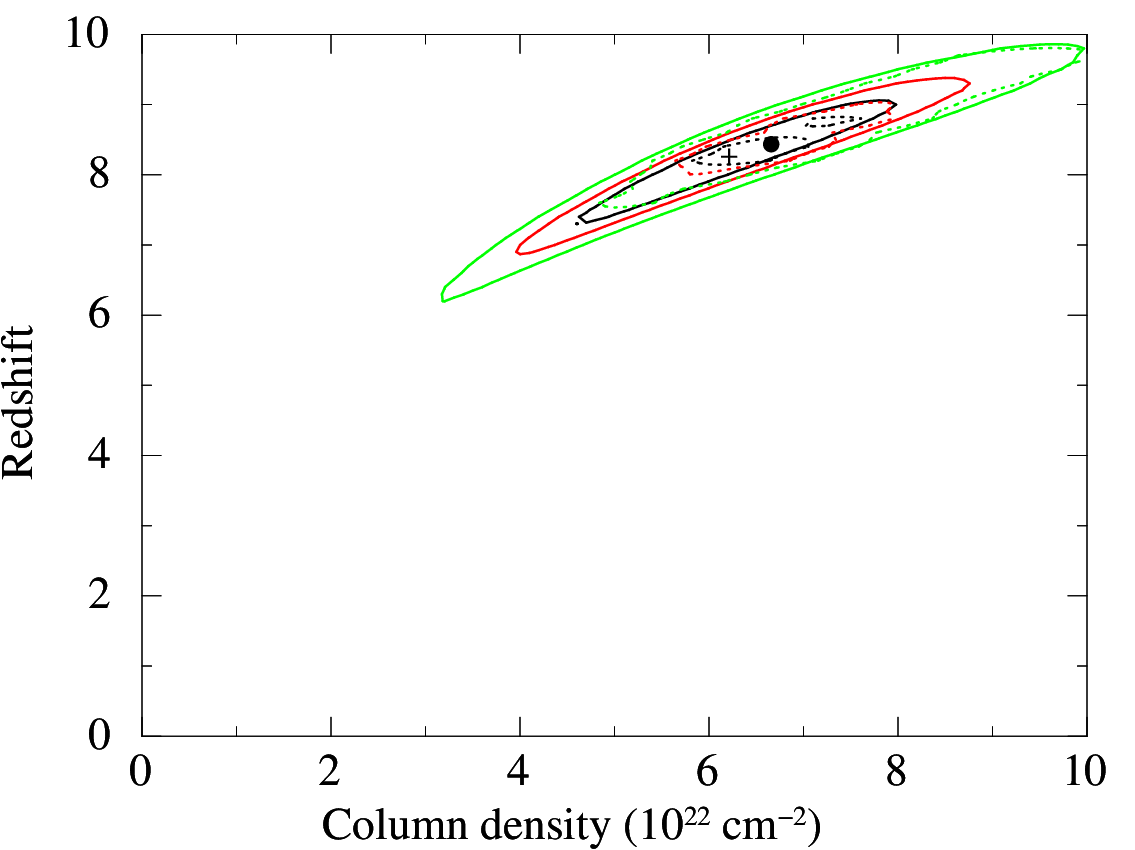}\\
\end{center}
\caption{Contour plot of the redshift vs. intrinsic column density based on simulated X--ray data for 
GRB050904, GRB080913, and GRB090423. Solid (dotted) contour lines represent 1, 2 and 3 $\sigma$ confidence level
for CCD (microcalorimeter) data. The cross (filled circle) the best fit value with CCD (microcalorimeter) data. }
\label{cont}
\end{figure}

Our starting point is the spectral fits provided by the Swift/XRT GRB spectrum repository (Evans et al. 2009).
These data were used as input values to simulate spectra (see Table 1). The Swift observations were taken as a 
template of the observational strategy that can be carried out by a low Earth orbiting satellite.
Therefore, the simulated observations start and last as do the Swift ones.
The adopted spectral model is a composite absorption modelled with {\tt phabs} (fixed to the Galactic value of 
Kalberla et al. 2005) and {\tt zphabs} plus a power law, within the spectral fitting program XSPEC.

As a first goal we try to recover the redshift of the GRB directly from the X--ray data.
This is clearly not needed for the bursts under consideration (because we know already their redshift) but
it is a useful exercise for GRBs that will have no optical/nIR redshift\footnote{On EXIST there will be an optical-infrared 
1.1 m telescope that will provide directly the redshift of distant
GRBs if not heavily obscured by dust absorption, as expected for
high-z GRBs.}.
Given these data, we cannot recover the redshift of GRB080913 due to its very low fluence (i.e. number of counts) 
but this can be easily done for GRB090423 and GRB050904 (see Table 2).

\begin{table}
\caption{Redshifts recovered from simulated X--ray observations.}
\begin{tabular}{ccccc}
\hline
GRB    & Simulated &True & $z$ limits & $z$ limits  \\
             &  counts$^+$       & $z$  & CCD obs.   & microc. obs.\\ 
\hline
050904 & 81397 (WT) & 6.29 & 5.9--6.8   & 6.0--6.4	 \\
050904&  78500 (PC) &          &                  &    \\
080913 & 3593  & 6.7  & 1.7--10    & 5.8--10     \\
090423 & 12776& 8.1  & 6.8--9.4   & 8.0--9.0    \\
\hline
080319B& $5.2\times 10^6$ & 7.0 &2.7--10.7 &3.3--8.7 \\
080319B$^*$& $4.8\times 10^6$ & 7.0& 6.5--7.1 & 6.9--7.0\\
\hline
\end{tabular}

$^+$ Simulated counts are indicative being random uncertainties included
in the simulations.

$^*$ artificially increased ($10^{22}\cmdue$) column density.
\end{table}

The next scientific question is to provide some indications on the metallicity of the absorbing medium.
Given the relatively low number of counts, this cannot be done on an element-by-element basis but by assuming 
a given distribution of metal abundances (e.g. solar composition), one can derive a global metallicity constraint 
(even without knowing how much hydrogen is present).
This search can however be carried out only after that a firm determination of the GRB redshift through 
optical/nIR spectroscopy has been gathered. 
We simulated then the same Swift XRT spectrum with the possibility of varying the metallicity, 
using the {\tt zvphabs} model within XSPEC.
When fitting a X--ray spectrum with an absorption model, we determine the column density 
with the implicit assumption of solar abundance ratios for the different elements versus hydrogen. 
In reality through X--ray fits, we determine the ratio of the column density over the metallicity, $Z$ (in solar units), as $N_H/Z$.
A first general limit on the metallicity of the absorbing medium can be derived assuming that the medium is Thomson thin. 
This means that $N_H/Z<1/\sigma_T=1.5\times 10^{24}\cmdue$ (where $\sigma_T$ is the Thomson scattering cross section). 
Therefore deriving a column density value naturally implies a lower limit on the metallicity, 
even for an uncertain neutral H column density (apart from the assumption of a solar composition). 
These numbers are given in Table 3. 

With a very large number of counts (see Table 2) one can try to break the degeneracy between metallicity and 
column density directly with the fitting procedure (leaving free the column density and the metallicity).
These results are reported in Table 3. The limits obtained with CCD observations are comparable to the ones obtained with the
Thomson limit assumption (i.e. from the determination of the column density and the assumption that the material is not Thomson thick). 
Those derived with the microcalorimeter are better only if the number of counts is high enough
(i.e. for GRB050904).

As a best case we considered one of the brightest GRB observed by Swift, i.e. the naked-eye GRB080319B (Racusin et al. 2008).
This burst is not unique rivaling the recent bright bursts discovered by Fermi in brightness (e.g. GRB080916C or GRB090323,
Abdo et al. 2009).
GRB 080319B was observed by the Swift narrow field instruments 64 s after its discovery and its brightness made it
possible to follow it continuously with the BAT and XRT
instruments. This burst was relatively close (redshift $z=0.94$) and low
absorbed ($N_H=1.2\times 10^{21}\cmdue$). We consider a test-case with
GRB080319B exploding at $z=7$ either with its observed
column density or with an intrinsic column density of $10^{22}\cmdue$. 
Concerning the redshift determination, the low column density case is not favorable whereas the high column density case results
in a very precise redshift location of the GRB (see Table 2). In fact, despite the large number of counts, if the column density is relatively 
low ($N_H\lsim 10^{21}\cmdue$) it cannot imprint measurable signatures on the spectrum. The same occurs for the study of the 
metallicity. We are able to set only loose limits on the metallicity in the low column density case, but still at least one order of magnitude
better than the Thomson limit, and we are able to set tight limits on
the metallicity (within a factor of 3 of the input value) in the case
of high column density.

As a further note we point out that for very distant objects there is
the possibility that a fraction or even all of the 
observed column density is not due to material within the GRB host galaxy, but comes from  
intervening system(s) (Campana et al. 2006, 2009).
Based on quasar studies (Wolfe et al. 2005; P\'eroux et al. 2003) we simulated for each of
our bursts a distribution of line-of-sights (10,000 trials), evaluating the corresponding X--ray absorption 
at the GRB redshift scaling the intervening column density as it would be observed at the GRB redshift (i.e. as $(1+z)^{2.6}$,
using the absorption energy dependence of Morrison \& McCammon 1983).

This will provide a distribution of column densities for each GRB due to intervening systems only. 
It can be shown that for the three bursts under consideration the mode ($50\%$) of this 
distribution is below the $N_H$ observed value, but not the $90\%$ value.
In other words, there are line of sights where the contribution from the intervening systems  
dominate but they are a minority.
Additional evidence can be gathered to circumvent this problem such as studies of column density variability 
(see also below) or observations in the radio band from which one can probe the local density from 
optically thick synchrotron emission. In the case of GRB050904, Frail et al. (2006) estimated a very high 
local density ($n\sim 84-680$ cm$^{-3}$) whereas for GRB090423 Chandra et al. (2009) estimated $n\sim 1$ cm$^{-3}$.
This might indicate a sizable contamination from intervening systems in the case of GRB090423.

\begin{table}
\caption{Metallicity recovering from simulated X--ray observations (the input value is solar metallicity).}
\begin{tabular}{cccc}
\hline
GRB    &$Z/Z_\odot$ Thomson& $Z/Z_\odot$ limits & $Z/Z_\odot$ limits  \\
       & limit     & CCD obs.   & microc. obs.\\ 
\hline
050904 & $>0.03$   & $>0.04$ &$>0.13$ \\
080913 & $>0.02$   & --      &$>0.001$ \\
090423 & $>0.04$   & $>0.02$ &$>0.01$ \\
\hline
080319B   @ $z=7$      &$>8\times 10^{-4}$ & $>4\times 10^{-3}$ & $>2\times 10^{-2}$\\
080319B$^*$ @ $z=7$& $>0.007$                & $>0.3$                      & $>0.3$ \\
\hline
\end{tabular}

$^*$ artificially increased ($10^{22}\cmdue$) column density.

\end{table}

Robust evidence for a decrease with time of the X--ray absorbing column in GRB050904 was reported by 
several authors (Watson et al. 2006; Campana et al. 2007; Cusumano et al. 2007; Gendre et al. 2007). 
Campana et al. (2007) modelled the evolution of the column density due to the flash ionization
of the GRB and early afterglow photons, allowing constraints on the geometry of the absorbing cloud.
This was done by dividing the X--ray light curve into four time bins and evaluating  
the absorbed spectrum with a cut-off power law model (see Campana et
al. 2007 for more details).  A decrease of the column density
indicates the presence of a sizable amount of matter close to the GRB site.
We simulate also this case obtaining a factor of $\sim 6$ ($\sim 8$ with the microcalorimeter) reduction in the size of the 
column density error bars or alternatively a factor of $\sim 10$ improvement in the number of column density bins in the 
temporal evolution with the same Swift XRT error bar sizes. These studies cannot be carried out with GRB080913 
and GRB090423 due to their relative faintness.

\section{Observability of Population III stars as GRBs}

Based on simulations it is thought that the first stars (Population
III) with typical masses of hundreds of solar masses 
formed in dark matter minihaloes around $z\sim 20-30$  (e.g. Bromm et al. 2009). The direct detection of single massive Pop. III 
stars is beyond the capabilities even of the upcoming James Webb Space Telescope (JWST), unless they explode 
as energetic pair-instability supernovae (PISNs).
PISNs would be bright enough to be readily detectable with the JWST at any relevant redshift (Scannapieco et al. 2005) 
but their number
density would be very low ($\sim 4$ deg$^2$ yr $^{-1}$, Weinmann \& Lilly 2005). It is not yet clear if Pop. III stars 
can give rise to GRBs but it is likely that these explosions will be
extremely powerful (e.g. Fryer et al. 2001; Komissarov \& Barkov 2010, 
see also M\'esz\'aros \& Rees 2010) and detectable with future facilities (Salvaterra et al. 2010). 
Several studies have investigated the build up of the first HII regions around Pop. III stars. These studies indicate
that at the end of the star's lifetime stellar photo-heating has affected the inner $\sim 50$ pc producing the build-up 
of material at the edge of this cavity (e.g. Greif et al. 2009). The integrated column density is however low (a few $10^{20}\cmdue$)
in all the studies on this subject (e.g. Greif et al. 2009; Abel, Wise \& Bryan 2007; Yoshida et al. 2007; Alvarez, Bromm \& Shapiro
2006). The prospects of detecting the traces of this absorption in the X--ray spectrum of a GRB are very poor. 

The second generation of stars, already enriched above the so-called critical metallicity of $10^{-4}\,Z_\odot$ (Schneider et al. 2002, 
2003), form in larger gas mass haloes. The gas in these larger haloes then collapses into several 
clouds which fragment into multiple clumps, each of which becomes a star-forming core (Inoue, Omukai \& Ciardi 2007). 
In these conditions optical depths are larger, especially in X--rays thanks to the presence of the first metals. Clearly, depending 
on the position of the exploding star in the cloud one can observe a distribution of absorbing column densities, similar to the 
one in the molecular clouds of our Galaxy (Reichart \& Price 2002).  We can then reverse the argument. If we will observe a bright 
GRB at very high redshift ($z>6$) with no intrinsic absorption (and free of intervening systems) 
it will be a good candidate for having been produced by a Pop. III star. 
We can then estimate the upper limit on the column density we can set in the case of a GRB050904-like burst with no intrinsic 
absorption. In the case of CCD or microcalorimeter observations we can set a $90\%$ confidence level upper limit 
of $\sim 5\times 10^{20}\cmdue$ on the intrinsic column density at $z=6.29$.

\section {Discussion and conclusions}

The discovery of GRB090423 has opened a completely new window on the high redshift universe.
It testifies that massive stars were already in place when the
Universe was only $\lsim 1$ Gyr old and that these can explode as GRBs.
These GRBs represent a unique opportunity to probe the circumburst medium and the ISM at these epochs. 
Due to the lack of bright high redshift quasars this is the only way we have to probe the ISM at 
such high redshifts. 

Starting from existing Swift XRT observations of very high redshift GRBs ($z>6$, GRB050904, GRB080913 
and GRB090423), we have simulated what could in principle be observed with a X--ray telescope with the same 
fast slewing capabilities of Swift and with a factor of $\sim 10$ larger effective area. We considered also two different
focal plane detectors: a CCD and a microcalorimeter. These characteristics are matched by the proposed X--ray telescopes
on board the EXIST (Grindlay 2009, 2010) and Xenia (Kouveliotou \& Piro 2008; Piro et al. 2010)
missions. These instruments (particularly EXIST) will also increase considerably 
the number of detected high redshift GRBs.

The first goal is the determination of the redshift based only on X--ray data. 
This can be extremely important if no optical/nIR observations are available or if the optical/nIR afterglow is too 
faint to obtain a spectroscopic redshift. Note that X-ray (only)
redshifts can also pre-select very high redshift candidates.
At these redshifts, the most prominent edges are those of S (0.35 keV at $z=6$ and 
0.27 keV at $z=8$) and Fe (1.01 keV at $z=6$ and 0.79 keV at $z=8$). Smaller edges come from  
Ar, Ca and Ni. We note in particular that, even at such high redshifts, iron should have already been formed 
and dispersed in the ISM in the case of Population II progenitors that
were enriched by Population III PISNe 
explosions (Heger \& Woosley 2002).
Based on simulated data, the redshift of GRB090423 and GRB050904 can be recovered directly from the X--ray data 
with a $\sim 10\%$ accuracy (better results are obtained with
microcalorimeter observations), but not for GRB080913
due to its faintness.
This points out the existence of a number of counts (or fluence) threshold, which depends also on the intrinsic 
(as well as Galactic) column density and on the GRB redshift. 
We also tested similar conditions of high intrinsic absorbing column density (a few $10^{22}\cmdue$), 
low Galactic column density (a few $10^{20}\cmdue$) as in the three cases considered here, finding a 
consistent fluence threshold in the $(3-4)\times 10^{-8}$ erg $\cmdue$ range. 
Provided that the observed spectrum is not varying, photons can be
collected over any time interval to accumulate this fluence. 
Given the decay of the afterglow however, to collect more photons it is more important to have a faster slew 
than a relatively large improvement in the telescope effective area. 

The determination of the metallicities and of column density variations is even more demanding in terms 
of photons. Only in the case of GRB050904 can we derive directly from X--ray data an estimate of the circumburst medium 
metallicity. A rough fluence threshold of a few $10^{-7}$ erg $\cmdue$ can be set for this kind of study.
Given the spectral variability observed in GRB050904, the fitting procedure has to be carried out with care
(e.g., taking time intervals in which the X--ray colour does not vary) and
for curved spectral models (e.g. by fitting column densities with a cut-off power law, the inaccuracies due to spectral changes 
are enormously reduced). If X--ray flares are present they also need to be modelled properly. 

Extrapolating the observed spectrum of a bright local ($z\sim 1$) burst, such as GRB080319B, to high redshift ($z=7$) clearly 
indicates that bright and moderately absorbed ($N_H\sim 10^{22}\cmdue$) bursts will represent the best tools to investigate 
the circumburst medium, allowing for a precise and direct redshift determination using X--ray data only and to a fair estimate 
of the mean metallicity of the absorbing medium.

In addition, simultaneous or nearly simultaneous optical/nIR and X--ray studies can shed light on the properties of the 
dust in the circumburst medium at high redshift. This can be done with spectral energy distribution studies and its variations
with time or with dedicated spectroscopic studies, by searching for column density and dust-induced spectral variations.


We also investigated the prospects of observing X--ray signatures for the presence of massive Pop. III stars. We were able to provide 
only a `negative' test. The environment surrounding Pop. III stars is too poor in metals and too tenuous to produce
absorption signatures in the X--ray spectra of associated GRBs at the moment of the explosion. For this reason if we have 
a bright GRB at high redshift ($z>6$) with no intrinsic absorption this can be a good candidate for being produced 
by a Pop. III progenitor. Observations in other bands are however needed to prove this hypothesis.

In conclusion, X--ray spectral studies can be potentially extremely useful in pinpointing 
high redshift GRBs (especially those without optical/nIR emission), and if corroborated by further evidence, they 
can provide unique information on the circumburst medium in the early Universe.

\section*{Acknowledgments}
This work has been partially supported by ASI grant I/011/07/0.
This work made use of data supplied by the UK Swift Science Data Centre at the University of Leicester.

{}

\end{document}